\documentclass[pre,preprintnumbers,amsmath,amssymb]{revtex4}
\usepackage{amssymb}
\usepackage{latexsym}
\usepackage{epsfig}
\begin{document}
\title{Thermodynamics of flat FLRW universe in Rastall Theory}
\author{H. Moradpour\footnote{h.moradpour@riaam.ac.ir}}
\address{Research Institute for Astronomy and Astrophysics of Maragha
(RIAAM), P.O. Box 55134-441, Maragha, Iran}

\begin{abstract}
In this paper, after referring to the Rastall theory, we address
some of its cosmological consequences. Moreover, bearing the
Clausius relation in mind, using Friedman equations in Rastall
theory and the Cai-Kim temperature, we obtain a relation for the
apparent horizon entropy of a flat FLRW universe. In addition, we
impose the entropy positivity condition on the obtained relation for
the horizon entropy, to find some constraints on the Rastall
parameters. Moreover, we investigate the second and generalized
second laws of thermodynamics. The results of considering a
dominated prefect fluid of constant state parameter are also
addressed helping us familiarize with the Rastall theory.
\end{abstract}
\maketitle

\section{Introduction}
Generalization of the $T^{\mu \nu}_{\ \ ;\mu}=0$ condition from flat
to the curved spacetimes is one of the Einstein's basic assumptions
to get the general relativity \cite{pois,gelen}. Indeed, by
insisting on this generalization in his attempt to formulate the
Mach principle, Einstein could get his famous tensor and thus the
corresponding field equations leading to the second order equations
of motion \cite{pois,gelen}, which have also vast applications in
the cosmological and astrophysical studies \cite{gelen,roos}. Many
years after Einstein, Jacobson showed that one can reobtain Einstein
equations by applying the Clausius relation on the local Rindler
causal horizon \cite{J1}. In fact, the Jacobson work proposes that
for spacetimes with a causal horizon the Einstein equations on the
horizon may be considered as a thermodynamical equation of state if
one generalizes the four law of black holes to the causal horizon,
i. e. a causal horizon may be taken into account as a proper causal
boundary. Moreover, the Jacobson's idea has been generalized to
$f(R)$ theory by Eling et al. showing that terms other than the
Einstein-Hilbert action produce entropy due to their non-equilibrium
thermodynamical aspects \cite{J11}, which leads to the modification
of the event horizon entropy \cite{J11,Caimore2}. One can also use
the Eling et al's proposal to get the equation of motion in
scalar-tensor gravity theory \cite{rep1}. It is useful to note here
that one may use the thermodynamics laws and gravitational field
equations to get the horizon entropy in the vast theories of gravity
\cite{T11}, which may satisfy the second law of thermodynamics
\cite{haw}. The horizon entropy, indeed, is not the total entropy in
a gravitational system. In fact, the total entropy of a
gravitational system, including the sum of horizon entropy and the
entropy of fields confined by the horizon, should increase during
every gravitational process. The latter referred to as the
generalized second law of thermodynamics \cite{GSL,GSL0}.

Introducing the unified first law of thermodynamics, Hayward could
show that the Einstein field equations on the trapping horizon of a
dynamic black hole are nothing but the unified first law of
thermodynamics \cite{Hay,Hay0,Hay2,Hay22}. This shows that the
trapping horizon may be considered as a causal boundary for
non-stationary spherically symmetric spacetimes. In fact, extending
the Hayward method to the apparent horizon of a FLRW universe, as
its causal boundary \cite{Bak}, we can get the apparent horizon
entropy and Friedman equations in the various theories of gravity
\cite{rep1,rep2,Cai2,Caimore2}. This approach may also be employed
to investigate the thermodynamic properties of apparent horizon of
the FLRW universe in some braneworld models such as the Gauss-Bonnet
and warped DGP braneworlds \cite{sheyw1,sheyw2}. Moreover, one can
also use the Hayward proposal to study the effects of interactions
between the dark energy and other parts of cosmos on the horizon
entropy in the Einstein and quasi-topological theories \cite{em,md}.
Here, it is worthwhile to mention that this approach does not lead
to the Friedman equations in the scalar-tensor theory, a result
which is due to the non-equilibrium thermodynamic aspect of the
scalar-tensor theory \cite{rep1}.

In another approach, Cai et al. applied the Clausius relation to the
apparent horizon of the FLRW universe, and used the horizon entropy
relation to get Friedman equations in various theories of gravity
\cite{CaiKim,rep2}. In addition, one can also use the Friedman
equations as well as the Clausius relation to get an expression for
the effects of interactions between the dark energy and other parts
of cosmos on the horizon entropy in the Einstein and
quasi-topological theories \cite{em,md,mm}. Although, following such
approaches, the definition of temperature differs from that of the
Hayward-Kodama temperature, the result for the horizon entropy is
equal to that of the approaches by which authors use the
Hayward-Kodama temperature and the unified first law of
thermodynamics \cite{CaiKim,Caimore1,em,md}. Besides, it seems that
the Cai-Kim temperature is more suitable than that of the
Hayward-Kodama for investigating the horizon thermodynamics
\cite{GSL1}. Indeed, their definition of temperature is equal to
that of a Hawking radiation for a locally defined apparent horizon
of the FLRW universe \cite{CaiKimt}, and may be used to investigate
the mutual relation between Friedman equations and the thermodynamic
properties of apparent horizon in various gravitational theories
\cite{CaiKim,rep2,Caimore1,em,md,mm,GSL1}. Here, it is useful to
note that the thermodynamic analysis of the total entropy of
universe, including the horizon entropy and the entropy of fields
confined by it, signals us to a universe satisfying thermodynamical
equilibrium conditions, in vast models of gravity \cite{msrw}.

In $1972$, by relating $T^{\mu \nu}_{\ \ ;\mu}$ to the derivative of
Ricci scalar, Rastall proposed a new formulation for gravity
\cite{rastall}, which converges to the Einstein formulation in the
flat background (empty universe). Indeed, he argued that the $T^{\mu
\nu}_{\ \ ;\mu}=0$ assumption, made by Einstein to obtain his field
equations, is questionable in the curved spacetimes \cite{rastall}.
In fact, the $T^{\mu \nu}_{\ \ ;\mu}\neq0$ condition is
phenomenologically confirmed by particle creation in cosmology
\cite{motiv1,motiv2,motiv3}. Indeed, in gravitational systems,
quantum effects lead to the violation of the classical condition
$T^{\mu \nu}_{\ \ ;\mu}=0$ \cite{motiv4}. Therefore, since $T^{\mu
\nu}_{\ \ ;\mu}$ is related to the Ricci scalar, the Rastall theory
may be considered as a classical formulation for the
particle creation in cosmology \cite{prd}, and it helps us in
investigating the possibility of coupling the geometry to the matter
fields in a non-minimal way. For the first time, Smalley tried to
get a lagrangian for this theory \cite{smal}. In addition, it seems
that astrophysical analysis, including the Neutron stars evolution,
and cosmological data do not reject this theory
\cite{neutrast,obs1,obs2}. Recently, this theory attracts more
investigators to itself, and its combinations with the Brans-Dicke
and Scalar-Tensor theories of gravity can be found in
\cite{rastbr,rastsc}. More investigations on the various aspects of
the Rastall theory in the context of current phase of the universe
expansion can also be found in
\cite{cosmos3,prd,rascos1,rasch,more1,more2,more3}. It is also shown
that this theory reproduces some loop quantum cosmological features
of the universe expansion \cite{more4}.

Our aim in this paper is to study the thermodynamics of the FLRW
universe in the Rastall theory. For this propose, after referring to
the Rastall theory, we address some of its cosmological features. In
addition, by using Friedmann equations in the Rastall theory,
attributing the Cai-Kim temperature to horizon and applying the
Clausius relation on the apparent horizon of the FLRW universe, we
get a relation for the apparent horizon entropy in the Rastall
theory. Since the entropy of a physical system is to be a positive
quantity \cite{CALLEN}, we study the effects of applying this
condition to the horizon entropy. The second and generalized second
laws of thermodynamics are also investigated. Our investigation
shows that the thermodynamic analysis of the FLRW universe imposes
some restrictions on the Rastall theory parameters which are in line
with previous studies. The results of considering a universe filled
by a prefect fluid with constant state parameter are also addressed.

The paper is organized as follows. In the next section, after
referring to the Rastall theory, we derive the corresponding
Friedman, Raychaudhuri and continuity equations and point to some of
the cosmological consequences of the Rastall theory. In addition,
some general remarks of the FLRW universe are also addressed.
Bearing the Cai-Kim temperature together with the Clausius relation
in mind, we use the Friedman and continuity equations to get the
horizon entropy in the Rastall theory, in section ($\textmd{III}$).
The results of imposing the entropy positivity condition on the
obtained relation for the horizon entropy are also studied. We also
investigate the second and generalized second laws of thermodynamics
in the third section. Throughout the paper, the results of
considering a prefect fluid with constant state parameter filling
the universe, are investigated in more details. Section
($\textmd{IV}$) is devoted to a summary and concluding remarks.
Throughout this paper we set $G=\hbar=c=1$ for the sake of
simplicity.

\section{Rastall theory and basic assumptions}
Rastall questioned the assumption $T^{\mu \nu}_{\ \ ;\mu}=0$ in
curved spacetime, and gets a new theory for gravity by proposing
$T^{\mu \nu}_{\ \ ;\mu}=\lambda R^{\ ,\nu}$, where $\lambda$ is an
unknown constant which should be specified from observations and
other parts of physics \cite{rastall}. Therefore, for a spacetime
metric $g_{\mu \nu}$, the corresponding gravitational field
equations can be written as
\begin{eqnarray}\label{r1}
G_{\mu \nu}+k\lambda g_{\mu \nu}R=kT_{\mu \nu},
\end{eqnarray}
where $G_{\mu \nu}$ and $T_{\mu \nu}$ are the Einstein and
energy-momentum tensors, respectively \cite{rastall}. Moreover, $R$
is Ricci scalar, and $k$ is also gravitational constant in Rastall
theory and should probably be specified from other parts of physics
and observations \cite{rastall}. It is obvious that for $\lambda=0$
and $k=8\pi$ the Einstein field equations are reobtained wherever
$T^{\mu \nu}_{\ \ ;\mu}=0$ \cite{rastall}.

Consider a cosmological background described by FLRW metric:
\begin{eqnarray}\label{frw}
ds^{2}=-dt^{2}+a^{2}\left( t\right) \left[ \frac{dr^{2}}{1-\kappa r^{2}}%
+r^{2}d\Omega^{2}\right].
\end{eqnarray}
Here, $a(t)$ and $\kappa$ denote the scale factor and the curvature
parameter, respectively, while $\kappa=-1,0,1$ denotes the open,
flat and closed universes, respectively \cite{roos}. Apparent
horizon as the marginally trapped surface of FLRW universe is
defined as
\begin{eqnarray}\label{ah2}
\partial_{\alpha}\tilde{r}\partial^{\alpha}\tilde{r}=0\rightarrow r_A,
\end{eqnarray}
where $\tilde{r}=a(t)r$, leading to
\begin{eqnarray}\label{ah}
\tilde{r}_A=a(t)r_A=\frac{1}{\sqrt{H^2+\frac{\kappa}{a(t)^2}}},
\end{eqnarray}
for the physical radii of apparent horizon. In fact, it seems that
the apparent horizon can play the role of causal boundary for the
FLRW spacetime \cite{Hay2,Hay22,Bak,sheyw1,sheyw2}. Since
cosmological data points to a flat universe \cite{roos}, we only
consider the flat case ($\kappa=0$) throughout this paper. For a
prefect fluid source ($T_{\mu}^{\nu}=\textmd{diag}(-\rho,p,p,p)$),
by using Eqs.~(\ref{r1}) and~(\ref{frw}), we get the corresponding
Friedmann equations in the Rastall theory
\begin{equation}\label{friedman1}
(12k\lambda-3)H^2+6k\lambda \dot{H}=-k\rho,
\end{equation}
and
\begin{equation}\label{friedman2}
(12k\lambda-3)H^2+(6k\lambda-2) \dot{H}=kp,
\end{equation}
where $\rho$ and $p$ are the energy density and pressure of the
energy-momentum source, respectively. The Rastall field equations
can also be written as
\begin{equation}\label{ein}
G_{\mu \nu}=\kappa (T_{\mu\nu}-\frac{\kappa\lambda
T}{4\kappa\lambda-1}g_{\mu\nu}),
\end{equation}
in which $T$ is the trace of energy-momentum tensor. Bianchi
identity implies $G_{\mu \nu}^{;\mu}=0$, which finally leads to
\cite{cosmos3}
\begin{equation}\label{cont}
(\frac{3k\lambda-1}{4k\lambda-1})\dot{\rho}+(\frac{3k\lambda}{4k\lambda-1})\dot{p}+3H(\rho+p)=0,
\end{equation}
as the continuity equation in the Rastall theory. It is worth
mentioning here that one can rediscover the Friedmann and continuity
equations in the Einstein theory by inserting $\lambda=0$ and
$k=8\pi$ in the above formulas. Moreover, by combining
Eqs.~(\ref{friedman1}) and~(\ref{friedman2}) with each other, one
may get the Raychaudhuri equation
\begin{equation}\label{rey}
\dot{H}=-\frac{k}{2}(\rho+p),
\end{equation}
which is independent of $\lambda$, and it is indeed the same as that
of the standard cosmology, which is based on the Einstein theory and
the FLRW metric. For a fluid with state parameter
$\omega=\frac{p}{\rho}$, by combining this equation with
Eq.~(\ref{friedman1}) one can get
\begin{equation}\label{nfried}
H^2=\frac{k(3k\lambda(1+\omega)-1)}{3(4k\lambda-1)}\rho.
\end{equation}
Moreover, for a fluid, the $\rho(a)$ relation may be obtained by
inserting $\omega=\frac{p}{\rho}$ into Eq.~(\ref{cont}) and taking
integral from the result. This leads to $\rho=\rho_0
a^{\frac{-3(1+\omega)(4k\lambda-1)}{3k\lambda(1+\omega)-1}}$ for a
fluid of constant state parameter. It is easy to check that the
famous first Friedman equation in the Einstein relativity framework
is reobtained by inserting $\lambda=0$ and $k=8\pi$ into
Eq.~(\ref{nfried}). Moreover, by combining Eqs.~(\ref{friedman1})
and~(\ref{friedman2}) one reaches
\begin{eqnarray}\label{1}
kT=2(12k\lambda-3)(2H^2+\dot{H}),
\end{eqnarray}
in which $T=3p-\rho$ is the trace of energy-momentum source. For a
traceless energy-momentum tensor, this equation leads to
$H=\frac{1}{2t}$ which is nothing but the Hubble parameter of
radiation dominated era \cite{roos}. Therefore, its prediction about
the Hubble parameter of the radiation dominated era is the same as
that of Einstein theory (the Friedman equations). In addition, for a
pressureless source ($p=0$), Eq.~(\ref{friedman1}) yields
$H=\frac{(6k\lambda-2)}{(12k\lambda-3)t}$ which, only for
$\lambda=0$, covers the Friedman results about the matter dominated
era ($H=\frac{2}{3t}$). Briefly, apart of a coefficient, this theory
predictions about the Hubble parameter of the matter dominated era
is the same as that of Einstein theory. Now, regarding a universe of
a constant density ($\rho=\rho_0$) in which $\omega=-1$ that leads
to a constant density, we can rewrite Eq.~(\ref{nfried}) as
\begin{equation}\label{nnfried}
H=\sqrt{\frac{-k}{3(4k\lambda-1)}\rho_0}.
\end{equation}
Therefore, for the $k>0$ case, a Rastall theory with
$\lambda<\frac{1}{4k}$ leads to a constant positive Hubble parameter
if $\rho_0>0$. Additionally, a source with $\rho_0<0$ may also lead
to a constant positive Hubble parameter whenever $k>0$ and
$\lambda>\frac{1}{4k}$. In sum, the Rastall theory with $k>0$ may
cover the primary inflationary era ($\omega=-1$ and $\rho_0>0$), if
$\lambda<\frac{1}{4k}$. In fact, since $H^2$ is a positive quantity
($H^2>0$), for the $k>0$ case, the RHS of Eq.~(\ref{nfried}) will be
positive for a fluid with $-1\leq\omega\leq\frac{1}{3}$ and
$\rho>0$, if $\lambda$ either satisfies $\lambda<\frac{1}{4k}$ or
$\lambda\geq\frac{1}{3k(1+\omega)}$. These results indicate that the
Rastall theory may also be used to describe the current accelerating
phase \cite{cosmos3,prd,rascos1,rasch,more1,more2,more3}, as well as
the primary inflationary era \cite{more4}.
\section{Thermodynamics of apparent horizon}
Since the Rastall gravitational field equations~(\ref{r1}) differ
from those of the Einstein theory, the Rastall lagrangian also
differs from that of Einstein \cite{smal}. Therefore, one may expect
that the horizon entropy in this theory differs from the Bekenstein
entropy. Now, by using the Cai-Kim approach \cite{CaiKim}, we try to
get an expression for the horizon entropy. Indeed, we apply the
Clausius relation on the apparent horizon and use the Cai-Kim
approach together with the Friedmann equations in the Rastall
theory, to get a relation for the horizon entropy in this theory.
Following that, we study the results of imposing the entropy
positivity condition on the obtained entropy relation. Finally, we
point to the second law of thermodynamics \cite{haw} and required
conditions for meeting this law in the corresponding Rastall
cosmology.
\subsection{The entropy of apparent horizon}
In the Cai-Kim approach, horizon temperature meets the
$T=\frac{H}{2\pi}$ relation in the flat FLRW universe, while the
volume change of universe in the infinitesimal time $dt$ is
neglected ($dV\approx0$) \cite{CaiKim,Cai2,CaiKimt}. The projection
of the total four-dimensional energy-momentum tensor $T^b_a$ in the
normal direction of the two-dimensional sphere with radii
$\tilde{r}$ is defined as \cite{CaiKim,Cai2,CaiKimt}
\begin{eqnarray}\label{esv}
\psi_a = T^b_a\partial_b \tilde{r} + W\partial_a \tilde{r},
\end{eqnarray}
where $W=\frac{\rho-p}{2}$ is the work density, and $a,b={t,r}$
\cite{CaiKim,Cai2,CaiKimt}. The energy flux ($\delta Q^m$) crossing
the apparent horizon during the infinitesimal time $dt$ and small
radius change $dr$ is defined as
\begin{eqnarray}\label{esv0}
\delta Q^m \equiv A\psi_a dx^a.
\end{eqnarray}
$A$ being the surface area of two-dimensional sphere with radii
$\tilde{r}$ \cite{CaiKim,Cai2,CaiKimt}. Simple calculations lead to
\cite{CaiKim,Cai2,CaiKimt,em,md,mm}
\begin{eqnarray}\label{ufl1}
\delta
Q^m=-\frac{3V(\rho+p)H}{2}dt+\frac{A(\rho+p)}{2}(d\tilde{r}-\tilde{r}
H dt).
\end{eqnarray}
In obtaining this equation we have used $d\tilde{r}=rda+adr$ and
$A\tilde{r}=3V$. By applying the $d\tilde{r} \approx 0$
approximation to this result, we get
\begin{eqnarray}\label{ufl3}
\delta Q^m=-3VH(\rho+p)dt.
\end{eqnarray}
Now, combining Eq.~(\ref{ufl3}) with the Clausius relation
($TdS_A=\delta Q^m$) and the Cai-Kim temperature
($T=\frac{H}{2\pi}$), we reach
\begin{eqnarray}\label{claus1}
dS_A\equiv -\frac{\delta Q^m}{T}=6\pi V(\rho+p)dt,
\end{eqnarray}
in which the extra mines sign comes from the universe expansion
\cite{CaiKim,Cai2,CaiKimt,em,md,mm}. Now, using Eq.~(\ref{cont}),
one can rewrite this equation as
\begin{eqnarray}\label{ent0}
dS_A=-\frac{8\pi^2}{3(4k\lambda-1)H^4}((3k\lambda-1)d\rho+(3k\lambda)dp).
\end{eqnarray}
Also, applying Eqs.~(\ref{friedman1}) and~(\ref{friedman2}) we get
\begin{equation}
d\rho=\frac{-1}{k}[2(12k\lambda-3)HdH+6k\lambda d\dot{H}],
\end{equation}
and
\begin{equation}
dp=\frac{1}{k}[2(12k\lambda-3)HdH+(6k\lambda-2)d\dot{H}],
\end{equation}
respectively. Inserting these two last equations into
Eq.~(\ref{ent0}) leads to
\begin{eqnarray}\label{ent1}
dS_A=-\frac{16\pi^2}{kH^3}dH,
\end{eqnarray}
also obtainable inserting Eq.~(\ref{rey}) into Eq.~(\ref{claus1}).
Taking integral from this equation and using the
$A=\frac{4\pi}{H^2}$ relation one obtains
\begin{eqnarray}\label{entf}
S_A=\frac{2\pi A}{k}.
\end{eqnarray}
This is nothing but the apparent horizon entropy in the Rastall
theory. It is easy to check that the Bekenstein limit (Einstein
result) is also deducible by inserting $k=8\pi$ parallel to the
$\lambda=0$ limit \cite{ahep}. Therefore, as authors in
Ref.~\cite{plb1c} properly address, the entropy of apparent
horizon of FRW universe in the Rastall theory meets
Eq.~(\ref{entf}), a result which is true for both the flat and
non-flat FRW universes \cite{plb1c}. On one hand, from
Eq.~(\ref{entf}), it seems that the Rastall parameter ($\lambda$)
does not affect the apparent horizon entropy \cite{plb1c}. One the
other hand, since the Rastall gravitational field equations and
its gravitational lagrangian differ from those of the Einstein
theory \cite{smal}, its horizon entropy must be different from the
horizon entropy in the Einstein theory \cite{ahep}. The key to
this puzzle lies in the Newtonian limit of the Rastall theory
implying that

\begin{eqnarray}\label{gamma1}
k=\frac{4\gamma-1}{6\gamma-1}8\pi,
\end{eqnarray}

\noindent with $\gamma=k\lambda$ as the Rastall dimensionless
parameter \cite{ahep}. Inserting this result into
Eq.~(\ref{entf}), one can obtain

\begin{eqnarray}\label{eins27b}
S=(1+\frac{2\gamma}{4\gamma-1})S_0,
\end{eqnarray}

\noindent where $S_0=\frac{A}{4}$ is the entropy of the apparent
horizon of FRW universe in the Esintein theory called the
Bekenstein entropy. It is apparent that for $\lambda=0$, parallel
to $\gamma=0$, the Esintein result ($S=S_0$) is covered.
Therefore, according to the Rastall hypothesis, the
$\frac{2\gamma}{4\gamma-1}S_0$ term is the modification to the
apparent horizon entropy of the FRW universe.

\subsection{The entropy positivity condition}
As we know, entropy is a positive quantity \cite{CALLEN}.
Regarding Eq.~(\ref{eins27b}), the $S\geq0$ condition implies that
the Rastall dimensionless parameter should either meet the
$\gamma>\frac{1}{4}$ or $\gamma\leq\frac{1}{6}$ conditions which
finally leads to
\begin{eqnarray}\label{la}
\lambda>\frac{1}{4k},\ \textmd{or}\ \lambda\leq\frac{1}{6k}
\end{eqnarray}
conditions for the $\lambda$ parameter, respectively. These
results are in full agreement with those obtained in \cite{ahep}.
\subsection{The second and generalized second laws of thermodynamics}
In order to investigate the second law of thermodynamics, we use
Eq.~(\ref{claus1}) to get
\begin{eqnarray}\label{claus2}
\frac{dS_A}{dt}=6\pi V(\rho+p),
\end{eqnarray}
i. e. the second law of thermodynamics ($\dot{S}_A\geq0$) is obeyed
if the prefect fluid, which supports the geometry, satisfies the
$\rho+p\geq0$ condition. Indeed, if we define the state parameter to
be $\omega=\frac{p}{\rho}$, so $\omega$ is not necessary constant, then for $\omega\geq-1$ and $\rho\geq0$ the second law
of thermodynamics is obtainable. Here, it is also useful to note that
this result is the same as those of the Einstein and
quasi-topological gravity theories \cite{em,md}. Finally, by
inserting Eq.~(\ref{rey}) into this equation, one obtains
\begin{eqnarray}
\frac{dS_A}{dt}=-\frac{12\pi V \dot{H}}{k}.
\end{eqnarray}
Since cosmological data indicates $\dot{H}\leq0$ \cite{roos}, the
second law of thermodynamics is met if $k>0$ which is in agreement
with the result of employing the entropy positivity condition. The
generalized second law of thermodynamics states that the sum of the
entropy of cosmos parts, which include the horizon and the fluids
confined, should increase during the universe expansion
\cite{GSL,GSL0}. In order to study this law, we need evaluate the
entropy of prefect fluid supporting the background. Taking into
account the apparent horizon as the boundary and using the Gibbs law
\cite{CALLEN} as
\begin{equation}\label{Gibbs1}
T_mdS_m=dE+pdV,
\end{equation}
where $S_m$ and $T_m$ are the matter entropy and temperature,
respectively, one obtains
\begin{equation}\label{Gibbs2}
T_mdS_m=(\rho+p)dV+Vd\rho.
\end{equation}
In obtaining this equation we used the $E=\rho V$ relation. From now
on, we assume that horizon and prefect fluid confined by it are in
thermal equilibrium i. e. $T_m=T=\frac{H}{2\pi}$. In fact, since,
due to their temperature difference, an energy flux between horizon
and the materials confined has not yet been observed, such
assumption is not far from reality. Moreover, it is shown that there
is a Hawking radiation with the Cai-Kim temperature for the fields
near the apparent horizon of FLRW universe \cite{CaiKimt}, and thus,
such Hawking radiation may be considered as a mechanism for
producing such assumed thermal equilibrium. Therefore, the
$T_m=\frac{H}{2\pi}$ assumption is not an unlikely guess. Now, by
combining this result with Eqs.~(\ref{claus2}) and~(\ref{Gibbs2})
and using Eq.~(\ref{rey}) together with the $V=\frac{4\pi}{3H^3}$
relation we obtain
\begin{eqnarray}\label{Gibbs3}
T(dS_A+dS_m)=3VH(\rho+p)(1+\frac{k}{2H^2})dt+Vd\rho,
\end{eqnarray}
yielding
\begin{equation}\label{total1}
\frac{dS_T}{dt}=6\pi V(\rho+p)(1+\frac{k}{2H^2})+\frac{2\pi
V}{H}\dot{\rho},
\end{equation}
where $S_T=S_A+S_m$ is the total entropy. Since the generalized
second law of thermodynamics states that $S_T$ should satisfy the
$\dot{S}_T\geq0$ condition \cite{GSL}, this law will be met if the
\begin{equation}\label{total}
-3(\rho+p)H(1+\frac{k}{2H^2})\leq\dot{\rho},
\end{equation}
condition is obeyed by the prefect fluid source. Bearing the results
obtained from investigating the second law in mind, it is obvious
that for $k>0$, $-1\leq\omega$ and whenever $\dot{\rho}$ meets
Eq.~(\ref{total}), the second law of thermodynamics and its
generalized form are satisfied simultaneously. Combining this
equation with Eq.~(\ref{cont}) we get
\begin{equation}\label{total2}
\dot{p}\leq \frac{\rho+p}{2H\lambda}(3k\lambda-2H^2\lambda-1),
\end{equation}
which expresses that the generalized second law of thermodynamics
holds if the time derivative of the prefect fluid pressure satisfies
this equation. Now, we focus on a dominated fluid of constant state parameter. Since for a prefect fluid with constant state
parameter, $\dot{p}=\omega\dot{\rho}$ (as is the case for
radiation), one can reach
\begin{equation}\label{omega}
\frac{1+2H^2\lambda-3k\lambda}{6H^2\lambda(1+\frac{k}{2H^2})}\leq\omega,
\end{equation}
by combining Eqs.~(\ref{total}) and~(\ref{total2}). Now, for a fluid
with $-1\leq\omega$, which satisfies the second law, we get
\begin{equation}\label{omega1}
\lambda\leq-\frac{1}{8H^2+6k},
\end{equation}
if
$\frac{1+2H^2\lambda-3k\lambda}{6H^2\lambda(1+\frac{k}{2H^2})}\leq-1$.
Moreover, if
$-1\leq\frac{1+2H^2\lambda-3k\lambda}{6H^2\lambda(1+\frac{k}{2H^2})}$
one reaches
\begin{equation}\label{omega2}
-\frac{1}{8H^2+6k}\leq\lambda.
\end{equation}
It is obvious that the results obtained in Eqs.~(\ref{omega1})
and~(\ref{omega2}) are in agreement with those obtained in
Eq.~(\ref{la}).
\section{Summary and concluding remarks}
After referring to the Rastall theory of gravity, we considered
the flat FLRW universe, and find the corresponding Friedman
equations in this theory. We also addressed some cosmological
consequences of this theory. Thereinafter, we used the Cai-Kim
approach to get the energy flux crossing horizon during the
infinitesimal time $dt$. In addition, by applying the Clausius
relation on the apparent horizon (as the causal boundary of
system) and using the obtained Friedman equations, we got a
relation for the apparent horizon entropy (Eq.~(\ref{entf})). We
saw that the entropy positivity condition restrict the possible
values of the $\gamma$ parameter and thus the $\lambda$ parameter.
Moreover, we found out that the validity of the second law of
thermodynamics requires that $k>0$ and $\rho+p\geq0$. The
generalized second law of thermodynamics has also been studied.
Our investigation shows that, in order to meet the generalized
second law of thermodynamics, the time derivatives of energy
density and pressure of source supporting the background should
satisfy a lower and an upper bound, respectively. Finally, for a
universe filled by a prefect fluid of constant state parameter,
the validity of the generalized second law has been addressed in
ample details.
\acknowledgments{The author is grateful to the anonymous reviewer for worthy hints and constructive comments.}

\end{document}